\documentclass{article}

\usepackage{arxiv}

\usepackage[utf8]{inputenc} 
\usepackage[T1]{fontenc}    
\usepackage{hyperref}       
\usepackage{url}            
\usepackage{booktabs}       
\usepackage{amsfonts}       
\usepackage{nicefrac}       
\usepackage{microtype}      
\usepackage{lipsum}
\usepackage{verbatim}
\usepackage{amsmath}

\usepackage{nomencl}
\usepackage{xcolor}

\usepackage{tcolorbox}
\usepackage{booktabs}
\usepackage{tikz}
\usepackage{tikzscale}
\usetikzlibrary{math}
\usepackage{enumitem}
\usetikzlibrary{external}
\usepackage{pgfplots}
\pgfplotsset{compat = newest}
 
\usepackage{float}
\usepackage{subcaption}

\usepackage{hyperref}

\makenomenclature

\usepackage{caption,graphicx,newfloat}

\DeclareFloatingEnvironment[
  fileext=lob,
  listname={List of Boxes},
  name=Box,
  placement=htp,
]{BOX} 

\newcommand\algoG{PEPDAS}
\newcommand\algoN{PEPGSM}

\title{Bandgap optimization in combinatorial graphs with tailored ground states: Application in Quantum annealing}

\author{
  Siddhartha Srivastava\thanks{\texttt{sidsriva@umich.edu}},  \space Veera Sundararaghavan\thanks{ \texttt{veeras@umich.edu}} \\
  Department of Aerospace Engineering\\
  University of Michigan\\
  Ann Arbor, MI 48109 \\ 
}

\begin{document}
\maketitle

\begin{abstract}
A mixed-integer linear programming (MILP) formulation is presented for parameter estimation of the Potts model. Two algorithms are developed; the first method estimates the parameters such that the set of ground states replicate the user-prescribed data set; the second method allows the user to prescribe the ground states multiplicity. In both instances, the optimization process ensures that the bandgap is maximized. Consequently, the model parameter efficiently describes the user data for a broad range of temperatures. This is useful in the development of energy-based graph models to be simulated on Quantum annealing hardware where the exact simulation temperature is unknown. Computationally, the memory requirement in this method grows exponentially with the graph size. Therefore, this method can only be practically applied to small graphs. Such applications include learning of small generative classifiers and spin-lattice model with energy described by Ising hamiltonian. Learning large data sets poses no extra cost to this method; however, applications involving the learning of high dimensional data are out of scope. 

\end{abstract}


\nomenclature{$G$}{Simple undirected weighted Graph}
\nomenclature{$\mathcal{V}$}{Set of Graph's vertices}
\nomenclature{$\mathcal{C}$}{Set of Graph's connection}

\nomenclature{$E$}{Potts Energy}
\nomenclature{$\boldsymbol{\theta}$}{Set of parameters for Potts model}
\nomenclature{$E_0$}{Energy of ground state}
\nomenclature{$E_1$}{Energy of $1^{st}$ excited state}
\nomenclature{$\Delta E$}{Band gap}

\nomenclature{$\mathcal{S}$}{Set of all possible states}
\nomenclature{$\mathcal{S}_G^{\boldsymbol{\theta}}$}{Set of all Ground states}
\nomenclature{$\mathcal{S}_E^{\boldsymbol{\theta}}$}{Set of all Excited states}
\nomenclature{$\mathcal{S}_D$}{Set of the Data states}

\nomenclature{$N_{TS}$}{Number of Total states}
\nomenclature{$N_{GS}$}{Number of Ground states}
\nomenclature{$N_{ES}$}{Number of Excited states}
\nomenclature{$N_{DS}$}{Number of Data states}
\nomenclature{$N_{L}$}{Number of Labels}
\nomenclature{$N_{V}$}{Number of Graph vertices}
\nomenclature{$N_{C}$}{Number of Graph connections}

\nomenclature{$\boldsymbol{S}$}{A state of the graph}
\nomenclature{$s_i$}{Label of Vertex with index $i$ }
\nomenclature{$v_i$}{Vertex with index $i$ }

\nomenclature{$p(\boldsymbol{S})$}{Probability of a state, $\boldsymbol{S}$}
\nomenclature{$\eta$}{Label of Vertex with index $i$ }

\printnomenclature

\section{Introduction}\label{sec:intro}
Potts energy model was initially developed to describe interacting spins on a crystalline lattice. Since then, it has become an archetypal model in other fields involving operations research, network theory, and physics of phase transition. The motion of biological cells was described by Graner and Glazier \cite{graner1992simulation} using a large-Q Potts model. A similar approach was used in \cite{matsci_mcpm} to study grain boundary motion in polycrystalline microstructures during thermally induced grain growth and recrystallization process. In such studies, the system's dynamics is represented as a transition probability governed by the model's energy description. These problems can be simulated using Monte Carlo based simulations. On the other hand, there are problems where the equilibrium solutions are required, for instance, in computer vision, Potts model is often used to describe the cut energy of a segmentation problem (c.f.\cite{boykov2001fast}). These problems are usually solved using the Graph-cut method. The computation of this process becomes exceedingly challenging as more generality is introduced. Bagon's thesis \cite{bagon2012discrete} provides an excellent review of these generalities and suggests practical algorithms.

\begin{figure}[tph]
\centering
\includegraphics[width=0.6\linewidth]{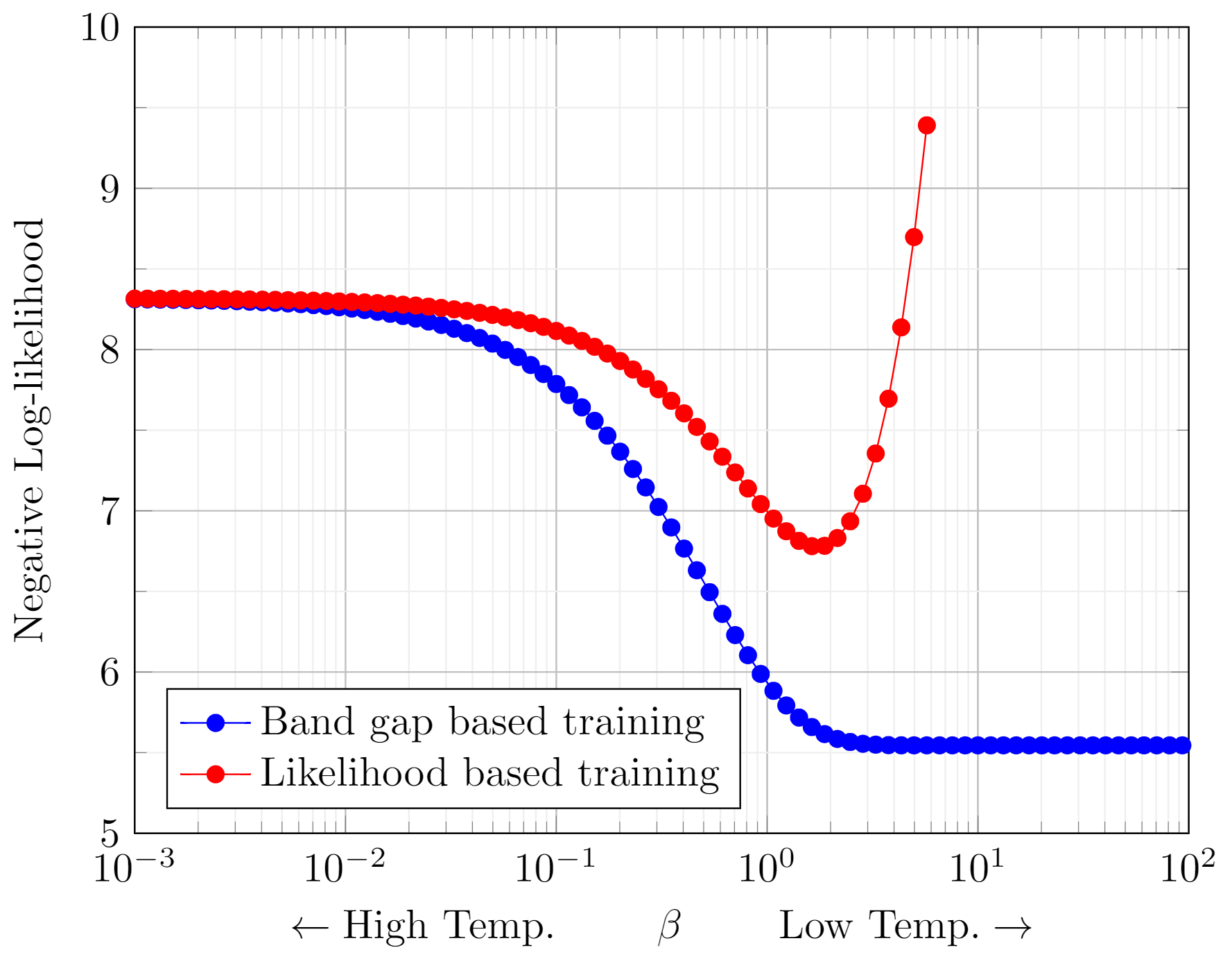}
\caption{Comparative analysis of likelihoods of models trained using the Likelihood maximization and band gap maximization. The predicted models are presented in Appendix \ref{app:K3}. A lower value of Negative Log likelihood signifies a better trained model}
\label{fig:NLLIllustration}
\end{figure}

Traditionally, these models are trained by considering them as Markov Random Fields (MRFs) and using gradient-based approaches to maximize the likelihood \cite{descombes1999estimation}. However, analytical estimates of the gradients are hard to compute. Among approximate techniques, Hinton's contrastive divergence method \cite{hinton2002training} provides an efficient way to approximate the gradients in the parameter optimization problem successively. An excellent review of this subject is presented in \cite{fischer2012introduction}. Recently, the advent of Quantum annealing technology has made it easier to sample states from the model's probability distribution \cite{adachi2015application}. This development has significantly eased the approximation of the required gradients. However, these methods have a critical drawback. These techniques only work for finite temperature probability distribution. Consequently, the model trained using these techniques is often temperature-dependent and shows disagreement with the data as the temperature is lowered \cite{srivastava2020machine}. As an example, the negative log-likelihood of a model trained using this technique is presented in Fig\ref{fig:NLLIllustration}. It can be seen that the minimum is close to the training $\beta$ (inverse temperature), which was chosen as $\beta=1$. A possible reason for this problem is that the training results in a locally optimal solution. Using quantum annealers adds another layer of complication because the simulation temperature is not known and depends on the graph size \cite{srivastava2020machine}. 

In contrast, this work is based on the band gap's maximization, while the ground states are chosen as the data states. This approach guarantees that the states' probability distribution gets closer to that of the data set as the temperature is reduced. Moreover, it ensures that the model adequately represents the data set for a broad range of temperatures. However, the downside of this approach is that there is no guarantee of the existence of parameters for every data set. This fact can be easily motivated by noticing that the number of ground states can be more than the number of model parameters and may result in an over-constrained optimization problem. Such problems do not exist at a non-zero temperature as all the states appear with non-zero probability. 

In this paper, a Mixed Integer Linear Programming (MILP) formulation is presented to estimate Potts model parameters. Two variations of the algorithm are presented. The first algorithm assigns a prescribed data set as the model's ground states while maximizing the bandgap. The second algorithm identifies a set of ground states with a prescribed multiplicity while maximizing bandgap. 
It should be noted that the computational complexity of both the algorithms grows exponentially with the size of the problem. Therefore, these methods are only suited for small graph structures. These problems arise in designing energies of smaller motifs in a lattice structure. 

The paper is organized as follows: The formulation for the Potts energy is reviewed in section \ref{sec:maths}. Concepts like the ground state, bandgap, and probability of a state are also reviewed. A theorem is presented to estimate the efficiency of the developed algorithms quantifiably. The problem statement is summarized in section \ref{sec:problem}. The developed algorithms are presented in section \ref{sec:method}. A case study for the Ising model is presented in \ref{sec:rnd}. Few details on the computational complexity are also outlined. Section \ref{sec:conclusion} provides a summary of the paper. 

\section{Mathematical Formulation}\label{sec:maths}

Potts model is a type of a discrete pairwise energy model on an undirected simple graph. In lieu of introducing some useful terms, following definition for graph is used:  

\textbf{Graph}: A graph, $G$, is a pair of sets $(\mathcal{V}, \mathcal{C})$, where $\mathcal{V}$ is the set of vertices and $\mathcal{C}$ is the set of
edges/connections. For each element $e\in \mathcal{C}$ there is a corresponding ordered pair $(x,y) ; x,y \in \mathcal{V}$ i.e. $\mathcal{C} \subseteq \mathcal{V}\times \mathcal{V}$. A Graph, $G=(\mathcal{V},\mathcal{C})$ is undirected if an edge does not have any directionality i.e $(x,y)\equiv (y,x)$. A graph is simple if $(x,x)\not\in \mathcal{C}$ for all $x\in \mathcal{V}$. 

Also, this work requires the graph to be finite, i.e., the number of vertices is finite. Next, the definition of Potts energy is introduced. 

\subsection{Potts model}
Consider a finite undirected simple graph $G(\mathcal{V},\mathcal{C})$. The number of vertices are denoted by $N_V=|\mathcal{V}|$ and the number of edges are denoted by $N_C=|\mathcal{C}|$. The indices of connections and vertices are related using the maps, $\pi_1$ and $\pi_2$ such that for a connection with index, $k\in \{1,..,N_C\}$, the index of the corresponding vertices are $\pi_1(k)$ and $ \pi_2(k)$ with $1\leq \pi_1(k) < \pi_2(k) \leq N_V$. This essentially means $e_k \equiv (v_{\pi_1(k)}, v_{\pi_2(k)})$. Each vertex, $v_i\in V$ is assigned a state $s_i \in \lbrace 1,2, \hdots , N_L\rbrace$ for all $i\in {1, \hdots ,N_V}$. This determines the complete state of the graph as an ordered tuple $\boldsymbol{S}=(s_1, \hdots ,s_i, \hdots , s_{n})\in \lbrace 1,\hdots,N_L \rbrace^{N_V}$. 
The set of all possible states is referred to as $\mathcal{S} =\lbrace 1, \hdots ,N_L \rbrace^{N_V}$ with the total number of states denoted by $N_{TS} = |\mathcal{S}| = N_L^{N_V}$. The Potts energy for a particular state can be evaluated as follows:
\begin{equation}\label{eq:PottsEnergy}
    E(\boldsymbol{S}) = \sum_{i=1}^{N_V} H_i U(s_i) + \sum_{k=1}^{N_C} J_{k} V\left(s_{\pi(k,1)},s_{\pi(k,2)}\right)
\end{equation}
where, $U(s)$ is the energy of labeling a vertex with label $s$, and $V(s_i,s_j)$ is the energy of labeling two connected vertices as $s_i$ and $s_j$. The parameters $H_i$ and $J_{k}$ are referred to as the Field strength and Interaction strength, respectively. 

Since the graph is undirected, following symmetry is imposed: 
\begin{align*}
    V(s_i, s_j) = V(s_j,s_i)
\end{align*}
The parameter set is represented as a vector, $\boldsymbol{\theta} =\begin{bmatrix}\theta_1, \hdots,\theta_{N_v + N_C}\end{bmatrix}^T$. In this work, it is specialized to following form: 
\begin{align*}
    \boldsymbol{\theta} = \begin{bmatrix}H_1, \hdots,H_{N_V},J_1,\hdots,J_{N_C}\end{bmatrix}^T
\end{align*}
This notation allows to describe energy as a matrix-product evaluated as $E(\boldsymbol{S}|\boldsymbol{\theta}) = \boldsymbol{\varepsilon}(\boldsymbol{S})\boldsymbol{\theta}$ where $\boldsymbol{\varepsilon}(\boldsymbol{S})$
\begin{align*}
    \boldsymbol{\varepsilon}(\boldsymbol{S}) = \begin{bmatrix}U(s_1), \hdots,U(s_{N_V}),V\left(s_{\pi_1(1)},s_{\pi_2(1)}\right),\hdots,V\left(s_{\pi_1(N_C)},s_{\pi_2(N_C)}\right)\end{bmatrix}
\end{align*}

\subsubsection{Ground states and band gap}

For a given set of parameters, $\boldsymbol{\theta}$, the set of ground states ($\mathcal{S}_G({\boldsymbol{\theta}}) \subseteq \mathcal{S}$) is the set of states with minimum energy, $E_{0}({\boldsymbol{\theta}})$), i.e. 
\begin{align*}
    \mathcal{S}_G({\boldsymbol{\theta}}) = \operatorname{argmin}_{\boldsymbol{S}\in \mathcal{S}} E(\boldsymbol{S}|\boldsymbol{\theta}), \qquad E_{0}({\boldsymbol{\theta}}) = \operatorname{min}_{\boldsymbol{S}\in \mathcal{S}} E(\boldsymbol{S}|\boldsymbol{\theta})
\end{align*}
In contrast, all the  non-minimal states are referred to as exited states. The set of all excited states, denoted by $\mathcal{S}_E({\boldsymbol{\theta}})$, can be evaluated as: 
\begin{align*}
    \mathcal{S}_E({\boldsymbol{\theta}}) = \mathcal{S}-\mathcal{S}_G({\boldsymbol{\theta}}) 
\end{align*}
The cardinalities of the set of ground states ($\mathcal{S}_G$) and excited states ($\mathcal{S}_E$) are denoted by $N_{GS}$ and $N_{ES}$, respectively.  All excited states may or may not have the same energy. However, the minimum excited energy referred to as the `first excited energy' is used in defining the band gap and is evaluated as:
\begin{align*}
    E_{1}({\boldsymbol{\theta}}) = \operatorname{min}_{\boldsymbol{S}\in \mathcal{S}_E({\boldsymbol{\theta}})} E(\boldsymbol{S}|\boldsymbol{\theta})
\end{align*}
 It should be noted that no assumption is made on the multiplicity of states with energy $E_1({\boldsymbol{\theta}})$. The band gap(a positive quantity) defines the energy gap between $\mathcal{S}_G$ and $\mathcal{S}_E$. It is estimated as: 
 \begin{align*}
     \Delta E (\boldsymbol{\theta}) = E_1(\boldsymbol{\theta}) - E_0(\boldsymbol{\theta} )
 \end{align*}

\subsubsection{Probability distribution}

At any given temperature, $T$, the probability of occurrence of a state, $\boldsymbol{S}$ is described by the Boltzmann distribution as: 
\begin{gather}\label{eq:boltzmann_prob}
    p(\boldsymbol{S}|\boldsymbol{\theta}, \beta) = \frac{1}{Z} e^{-\beta {E(\boldsymbol{S})}}
\end{gather}
where $\beta = 1/k_B T$ is the inverse thermodynamic temperature, $k_B$ is the Boltzmann constant and $Z$ denotes the partition function which is estimated as 
\begin{align*}
    Z = \sum_{\boldsymbol{S}\in \mathcal{S}} e^{-\beta{E(\boldsymbol{S})}} 
\end{align*}

\subsection{Parameter estimation}

Given a data set, $\mathcal{S}_D\subseteq \mathcal{S}$, the parameters set, $\boldsymbol{\theta}$, is optimized such that the states in $\mathcal{S}_D$ have higher probability of occurrence at a prescribed $\beta$ value. Mathematically, this procedure entails minimization of negative log-likelihood as defined below: 
\begin{gather}
    \eta ({\boldsymbol{\theta}, \beta}) = -\sum_{\boldsymbol{S}\in \mathcal{S}_D} \log{p(\boldsymbol{S} | \boldsymbol{\theta}, \beta)}
\end{gather}

It can be observed that at high temperatures i.e.  $\beta \rightarrow 0$, all states occur with equal likelihood and therefore 
\begin{align*}
    \eta_0 = \lim_{\beta\rightarrow 0} (\boldsymbol{\theta},\beta) = N_{DS} \log (N_{TS})
\end{align*}
where $N_{DS} = |\mathcal{S}_D|$. On the other hand, at low temperatures i.e. $ \beta \rightarrow \infty$, only ground states occur with equal probability and occurrence of any other state has probability 0. Consequently, the value of $\eta$ in this limit is finite only when $\mathcal{S_D} \subseteq \mathcal{S}_G $. It is evaluated as: 
\begin{align*}
    \eta_\infty  (\boldsymbol{\theta}) = \lim_{\beta\rightarrow \infty} (\boldsymbol{\theta},\beta) = N_{DS} \log(N_{GS})
\end{align*}

It is desirable to estimate parameters such that the ground state replicates the data set, and the bandgap is maximized. The reason will be apparent after the next theorem (proof in Appendix \ref{sec:theoremProof}). 

\textbf{Theorem}: For a given set of parameters, $\boldsymbol{\theta}_D$, such that (i) $\mathcal{S}_{G}(\boldsymbol{\theta}_D) = \mathcal{S}_D$ (ii) $\Delta E>0$, following statements hold true: 

\begin{enumerate}[label =(\alph*)]
    \item $\eta(\boldsymbol{\theta}_D,\beta)$ monotonically decreases with $\beta$ and the low temperature limit 
        \begin{align}\label{eq:lowTempEta}
            \eta_\infty  (\boldsymbol{\theta}_D)= \lim_{\beta\rightarrow \infty} \eta (\boldsymbol{\theta}_D,\beta) = N_{GS} \log(N_{GS}) 
        \end{align}
    \item $\eta(\boldsymbol{\theta}_D,\beta)$ is bounded as: 
\begin{align}
    N_{GS} \log(N_{GS}) < \eta(\boldsymbol{\theta}_D,\beta) \leq N_{GS} \log \left(N_{GS} + N_{ES} e^{-\beta \Delta E } \right)
\end{align}
    \item For any $\epsilon>0$, there exists a $\beta^*$ such that for all $\beta > \beta^*$,  $\eta (\boldsymbol{\theta}_D,\beta)  -  \eta_\infty (\boldsymbol{\theta}_D,\beta) < \epsilon $ where $\beta^*$ is estimated as: 
    \begin{align}\label{eq:betaMatch}
         \beta^* =  \frac{1}{\Delta E} \left(\log\frac{N_{ES}}{N_{GS}} - \log\left( e^{\epsilon /N_{GS}} - 1 \right)\right)
    \end{align}
\end{enumerate}

\begin{figure}[tph]
\captionsetup[subfigure]{labelformat=empty}
    \centering
    \begin{subfigure}[p]{0.48\textwidth}
    \centering
    \includegraphics[width=1\linewidth]{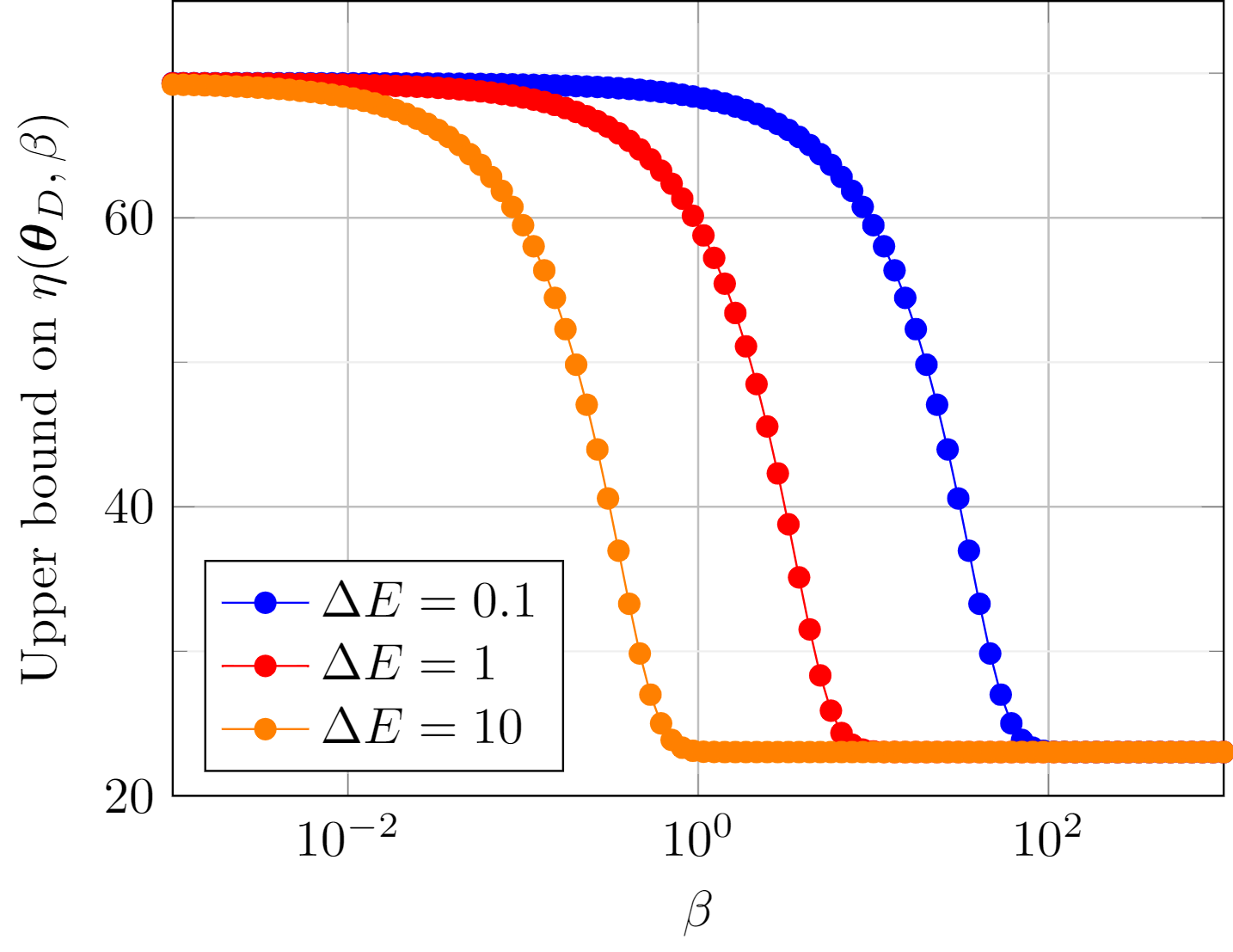}
    \caption{(a)}
    \end{subfigure}\hfill
    \begin{subfigure}[p]{0.48\textwidth}
    \centering
    \includegraphics[width=1\linewidth]{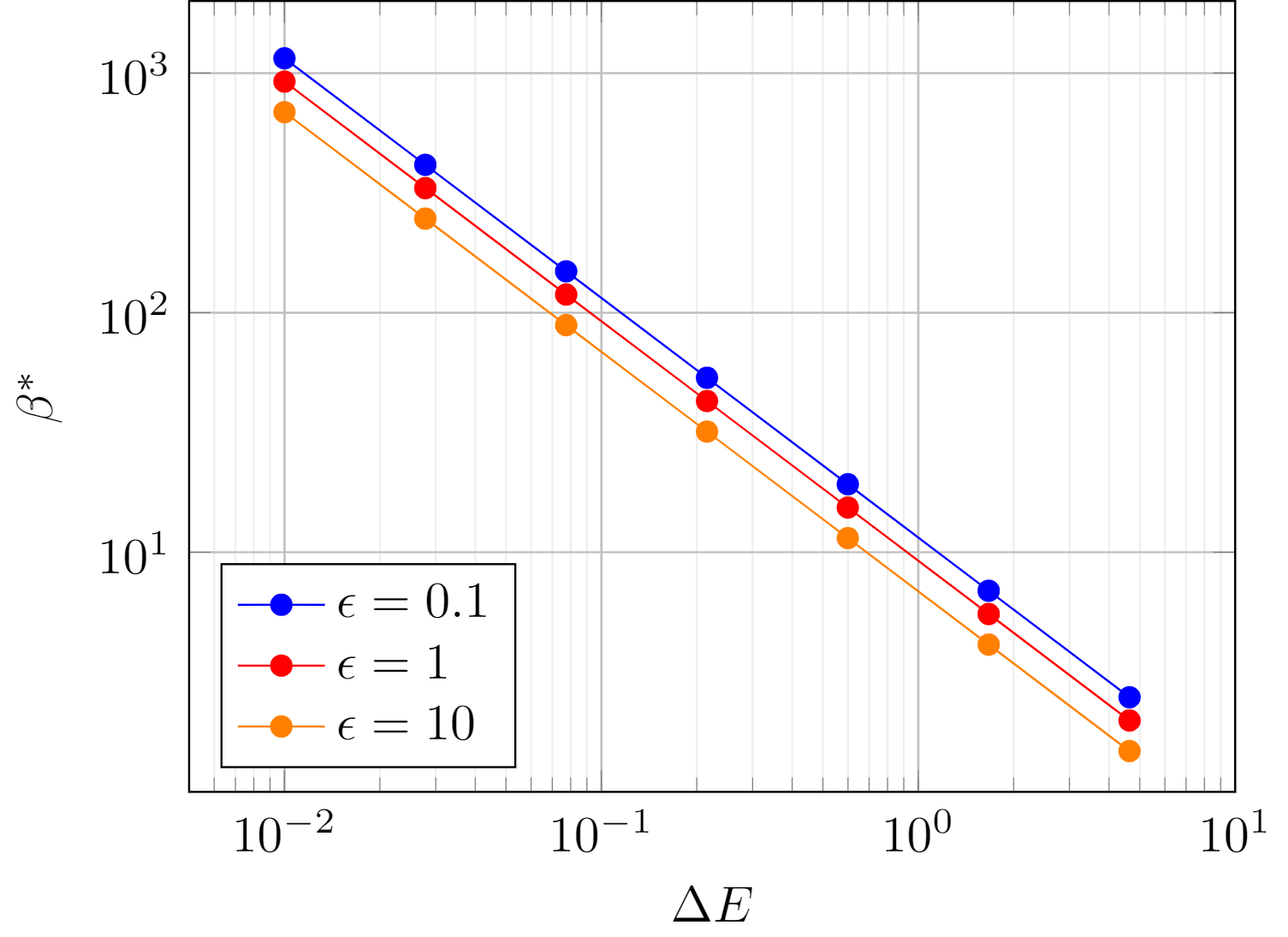}
    \caption{(b)}
    \end{subfigure}\hfill 
    \caption{An illustration of bounds for a trained Potts model with $N_{GS}=10$  and $N_{V} = 10$. (a) The upper bound on $\eta$ with respect to $\beta$ for various values of energy gap (b) $\beta^*$ as a function of band gap for various bounds on $\eta$}
\label{fig:BoundIllustration}
\end{figure}

The consequence of this theorem is that it guarantees that if the parameters are chosen appropriately, $\eta$ will approach to its global minimum in the low temperature (high $\beta$) limit. Moreover, at a finite $\beta$, $\eta$ is bounded from above by a decreasing function. It can be seen in Fig\ref{fig:BoundIllustration}(a), that the bound gets tighter for higher values of $\Delta E$. It is also shown that the trained model is efficient in the range of $\beta$ determined by $[\beta^*,\infty)$. Fig\ref{fig:BoundIllustration}(a) shows that a higher bandgap allows a broader range of temperatures.

\section{Problem Statement}\label{sec:problem}

Given a finite undirected simple graph $G(\mathcal{V},\mathcal{C})$, find parameters, $\boldsymbol{\theta}$ that maximizes the band gap in following two situations: 

\textit{Case 1}: $\mathcal{S}_D$ is prescribed and $\mathcal{S}_{G}(\boldsymbol{\theta}_D) = \mathcal{S}_D$.

\textit{Case 2}: Ground state multiplicity, $N_{GS}$, is prescribed.

To make this optimization problem well posed, it is additionally imposed that $H^{\min}_i\leq H_i\leq H^{\max}_i$ and $J^{\min}_{k}\leq J_{k}\leq J^{\max}_{k}$. Moreover, the functions $U(s)$ and $V(s_i,s_j)$
are predetermined and not calibrated in the optimization process.

\section{Methods}\label{sec:method}

A Mixed Integer Linear Programming (MILP) problem is formalized for parameter estimation of Potts model. A brief overview of the MILP formulation is presented below:  

\textbf{Mixed Integer Linear Programming (MILP)}:  An optimization problem is considered to be of MILP type when the objective function is linear in the decision variables and some of the decision variables are integer. A typical setup of MILP problem is given in Eq\eqref{eqn:MILP} where $\boldsymbol{x}$ is the decision variable of size $N$, $I$ is the set of indices of $\boldsymbol{x}$ which are integers and the matrices $\boldsymbol{A_{eq}}$, $\boldsymbol{b_{eq}}$, $\boldsymbol{A}$ and $\boldsymbol{B}$ are used to define linear constraints.
\begin{align}
\label{eqn:MILP}
\begin{split}
\text{Optimize:   }  &\quad \min_{\boldsymbol{x} } \boldsymbol{c}\boldsymbol{x} \\
\text{Inequality constraints:   } &\quad \boldsymbol{A}\boldsymbol{x} \leq \boldsymbol{b}\\
\text{Equality constraints:} &\quad \boldsymbol{A_{eq}} \textbf{x} = \textbf{b}_{eq}\\
\text{Bounds:   } &\quad \textbf{lb}\leq \textbf{x} \leq \textbf{ub} \\
 \text{Integer variables:    } &\quad \boldsymbol{x_I} \in \mathbb{Z}
\end{split}
\end{align}

The MILP formulation for the two cases is presented next. In both cases, the decision variables include the parameters, $\boldsymbol{\theta}$, and some auxiliary variables. These variables are introduced along with the algorithm description. Moreover, the algorithms do not enforce that $\Delta E >0$. Therefore, the results are accepted only if this condition is met.  

\subsection{Algorithm 1: Parameter Estimation for Potts model with DAta Set (\algoG{})}\label{sec:Algo1}

The energies of individual states can be evaluated as a matrix product operation (shown in Section \ref{sec:maths} ) which works well with linear programming framework. However, the calculation of band gap requires calculation of a minimum of energy over $\mathcal{S}_E$. This operation introduces a non-linearity. Thus, following auxiliary variables are introduced to pose this optimization as a linear programming problem: 

\begin{itemize}
    \item $E_1$ (real valued scalar): It represents the energy of the $1^{st}$ excited state.
    \item $\boldsymbol{m} = [m_1,...,m_{N_{ES}}]$ (binary valued vector of size $N_E$): It is defined such that it's value is 1 on exactly one index and 0 everywhere else. The index with value 1 must correspond to one of the $1^{st}$ excited state. 
    \item $M$ (real valued scalar): It represents a large positive number. For computational purposes it can be evaluated as: 
    \begin{align}
        M = \left( \max_{s} |U(s)| \right)  \sum_{i=1}^{N_V}  \left(|H_i^{\max}|+|H_i^{\min}|\right)  
        + \left( \max_{s_1,s_2} |V(s_1,s_2)| \right) \sum_{k=1}^{N_C}  \left(|J_k^{\max}|+|J_k^{\min}|\right) 
    \end{align}
\end{itemize}

The decision variable in this formulation are given as: 
\begin{align*}
    \boldsymbol{x} = \begin{bmatrix} 
    \boldsymbol{\theta}, & E_1, & \boldsymbol{m}
    \end{bmatrix}^T
\end{align*}

Consider a data set, $\mathcal{S}_D = \{\boldsymbol{\overline{S}}_1,..., \boldsymbol{\overline{S}}_{N_{DS}} \}$.  The optimization cost ($-\Delta E$) is estimated by substituting the $E(\boldsymbol{\overline{S}}_1)$ as that of ground state and $E_1$ for the 1st excited state energy. Thus the cost is evaluated as: 
\begin{align*}
    \text{Cost} = E(\boldsymbol{\overline{S}}_1) - E_1
\end{align*}
The energy of all data states are explicitly equated as follows: 
\begin{align*}
      E(\boldsymbol{S}_1) - E(\boldsymbol{S}_i)  & = 0  , \qquad \forall i\in\{2,...,N_{DS}\}
\end{align*}

The $1^{st}$ excited energy, $E_1$ is estimated by bounding it from above by energies of all the excited states. It is bounded from below by the energy of state corresponding to the index at which $m_i = 1$. The upper bound on $E_1$ insures that if $m_i=1$, then $E_1(\boldsymbol{\theta}) = E(\boldsymbol{S}_i)$. These conditions can be imposed using following set of equations and inequality: 
\begin{align*}
      E(\boldsymbol{S}_i) - E_1 + M m_i &\leq M  , \qquad \forall i\in\{1,...,N_{ES}\}\\
     -E(\boldsymbol{S}_i) + E_1         &\leq 0  , \qquad \forall i\in\{1,...,N_{ES}\}\\
     \sum_{i=1}^{N_{ES}} m_i &= 1 
\end{align*}

Most computing software only allows integer valued variables. In such a case, the binary value of variable $\boldsymbol{m}$ can be explicitly enforced by setting following bounds on integer valued $\boldsymbol{m}$: 
\begin{align*}
      0\leq m_i \leq 1 , \qquad \forall i\in\{1,...,N_{ES}\}
\end{align*}

This formulation is presented in Box 1 in the matrix format. 

\begin{BOX}[tph]\label{box:algo1}
\begin{tcolorbox}[
  opacityframe=1.0,
   colback=white,  
]
Optimization cost: 
\begin{align*}
    \boldsymbol{c} = \begin{bmatrix} 
    \boldsymbol{\varepsilon}(\boldsymbol{\overline{S}}_1) & -1 & \boldsymbol{0}_{1\times N_{ES}}
    \end{bmatrix}
\end{align*}
Inequality constraints
\begin{align*}
\boldsymbol{A} = \begin{bmatrix}
 \vdots& \vdots & \vdots\\
\boldsymbol{\varepsilon}(\boldsymbol{S}_{i}) & -1 & [0,...,0,\underbrace{M}_{i^{th}\text{index}},0,...,0]_{1\times N_{ES}}\\
-\boldsymbol{\varepsilon}(\boldsymbol{S}_{i}) & 1 & \boldsymbol{0}_{1\times N_{ES}}\\
 \vdots& \vdots & \vdots\\ 
\end{bmatrix} , \boldsymbol{b}= 
\begin{bmatrix}
\vdots\\
M\\ 
\\
0\\
\vdots 
\end{bmatrix}
\end{align*}

Equality constraints: 
\begin{align*}
\boldsymbol{A_{eq}} = \begin{bmatrix}
\boldsymbol{0}_{1\times N_V} & 0 & \boldsymbol{1}_{1\times N_{ES}}\\ 
\boldsymbol{\varepsilon}(\boldsymbol{\overline{S}}_2)- \boldsymbol{\varepsilon}(\boldsymbol{\overline{S}}_1) & 0 & \boldsymbol{0}_{1\times N_{ES}}\\ 
 \vdots& \vdots & \vdots\\ 
\boldsymbol{\varepsilon}(\boldsymbol{\overline{S}}_{N_{DS}})- \boldsymbol{\varepsilon}(\boldsymbol{\overline{S}}_1) & 0 & \boldsymbol{0}_{1\times N_{ES}}
\end{bmatrix} , \boldsymbol{b_{eq}}= 
\begin{bmatrix}
1  \\ 
0\\ 
 \vdots\\ 
0
\end{bmatrix}
\end{align*}

Bounds: 
\begin{align*}
    \boldsymbol{lb} &= \begin{bmatrix}
    H_1^{\min},...,H_{N_V}^{\min}, J_1^{\min},...,J_{N_C}^{\min}, -M, \boldsymbol{0}_{1\times N_{ES}}
    \end{bmatrix} \\
    \boldsymbol{ub} &= \begin{bmatrix}
    H_1^{\max},...,H_{N_V}^{\max}, J_1^{\max},...,J_{N_C}^{\max}, M, \boldsymbol{1}_{1\times N_{ES}}
    \end{bmatrix}
\end{align*}

\end{tcolorbox}
\caption{MILP formulation for \algoG{} method}
\end{BOX}

\subsection{Algorithm 2: Parameter Estimation for Potts model with Ground State Multiplicity (\algoN)}

In this formulation only the variable $N_{GS}$ is provided by the user in stead of $\mathcal{S}_Data$. This condition adds the complexity of locating the ground states and evaluating the ground state energy, $E_0(\boldsymbol{\theta})$. This problem is resolved by including following auxiliary variables: 
\begin{itemize}
    \item $E_0$ (real valued scalar): It represents the ground state energy.
    \item $\boldsymbol{l} = [l_1,...,l_{N_{TS}}]$ (binary valued vector of size $N_{TS}$): It is defined such that it's value is 1 on exactly $N_{GS}$ indices and 0 everywhere else. The index has value 1 if and only if it corresponds to the ground state. 
    \item $E_1$ and $M$ as defined in algorithm 1
    \item $\boldsymbol{m} = [m_1,...,m_{N_{TS}}]$ (binary valued vector of size $N_{TS}$): It is same as algorithm 1, except that the index are now enumerated based on the set $\mathcal{S}$
\end{itemize}

The decision variable in this formulation are given as: 
\begin{align*}
    \boldsymbol{x} = \begin{bmatrix} 
    \boldsymbol{\theta} & E_0 & E_1 & \boldsymbol{l} & \boldsymbol{m}
    \end{bmatrix} 
\end{align*}

The optimization cost is given as:  
\begin{align*}
    \text{Cost} = E_0 - E_1
\end{align*}

The estimation of $E_0$ is done using the same idea of bounding $E_0$ from above and below. The bound is tight only for indices where $l_i=1$. 
\begin{align*}
     -E(\boldsymbol{S}_i) + E_0         &\leq 0  , \qquad \forall i\in\{1,...,N_{TS}\}\\
      E(\boldsymbol{S}_i) - E_0 + M l_i &\leq M  , \qquad \forall i\in\{1,...,N_{TS}\}\\
     \sum_{i=1}^{N_{TS}} l_i &= N_{GS}
\end{align*}

For the estimation of $E_1$, the upper bound is lifted on indices corresponding to ground states. This allows to estimate minimum over non-optimal states. Moreover, index of $1^{st}$ excited state cannot coincide with ground state i.e. $l_i = 1$ and $m_i=1$ cannot occur simultaneously. These conditions are imposed using following inequalities and equations: 
\begin{align*}
     -E(\boldsymbol{S}_i) + E_1 - M l_i &\leq 0  , \qquad \forall i\in\{1,...,N_{TS}\}\\
      E(\boldsymbol{S}_i) - E_1 + M m_i &\leq M  , \qquad \forall i\in\{1,...,N_{TS}\}\\
      l_i + m_i &\leq 1 , \qquad \forall i\in\{1,...,N_{TS}\}\\
      \sum_{i=1}^{N_{TS}} l_i &= 1
\end{align*}

The condition of binary valued variables is imposed on integer variables as follows: 
\begin{align*}
      0\leq l_i, m_i \leq 1 , \qquad \forall i\in\{1,...,N_{ES}\}
\end{align*}

This formulation is presented in Box 2 in the matrix format. 

\begin{BOX}[tph]
\begin{tcolorbox}[
  opacityframe=1.0,
   colback=white,  
]
Optimization cost: 
\begin{align*}
    \boldsymbol{c} = \begin{bmatrix} 
    \boldsymbol{0}_{1\times (N_V + N_C)} & 1 & -1 & \boldsymbol{0}_{1\times N_{TS}} & \boldsymbol{0}_{1\times N_{TS}}
    \end{bmatrix}
\end{align*}
Inequality constraints
\begin{align*}
\boldsymbol{A} = \begin{bmatrix}
\boldsymbol{0}_{1\times (N_V + N_C)} & 0 & 0 & \boldsymbol{1}_{1\times N_{TS}} & \boldsymbol{1}_{1\times N_{TS}} \\
 \vdots & \vdots & \vdots & \vdots & \vdots\\
-\boldsymbol{\varepsilon}(\boldsymbol{S}_{i})   & 1     & 0    &  \boldsymbol{0}_{1\times N_{TS}}                                          &  \boldsymbol{0}_{1\times N_{TS}} \\
\\ 
\boldsymbol{\varepsilon}(\boldsymbol{S}_{i})    & -1    & 0    &  [0,...,0,\underbrace{M}_{i^{th}\text{index}},0,...,0]_{1\times N_{TS}}   &  \boldsymbol{0}_{1\times N_{TS}} \\
-\boldsymbol{\varepsilon}(\boldsymbol{S}_{i})   & 0     & 1    & [0,...,0,\underbrace{-M}_{i^{th}\text{index}},0,...,0]_{1\times N_{TS}}   &  \boldsymbol{0}_{1\times N_{TS}} \\
\boldsymbol{\varepsilon}(\boldsymbol{S}_{i})    & 0     &-1    &  \boldsymbol{0}_{1\times N_{TS}}                                          &  [0,...,0,\underbrace{M}_{i^{th}\text{index}},0,...,0]_{1\times N_{TS}} \\
 \vdots& \vdots & \vdots & \vdots & \vdots\\ 
\end{bmatrix} , \boldsymbol{b}= 
\begin{bmatrix}
1\\
\vdots\\
0\\ 
\\
M\\
\\
0\\
\\
M\\
\\
\vdots 
\end{bmatrix}
\end{align*}

Equality constraints: 
\begin{align*}
\boldsymbol{A_{eq}} = \begin{bmatrix}
\boldsymbol{0}_{1\times (N_V + N_C)} & 0 & 0 & \boldsymbol{1}_{1\times N_{TS}} & \boldsymbol{0}_{1\times N_{TS}} \\
\boldsymbol{0}_{1\times (N_V + N_C)} & 0 & 0 & \boldsymbol{0}_{1\times N_{TS}} & \boldsymbol{1}_{1\times N_{TS}} \\
\end{bmatrix} , \boldsymbol{b_{eq}} = 
\begin{bmatrix}
N_{GS}\\
1
\end{bmatrix}
\end{align*}

Bounds: 
\begin{align*}
    \boldsymbol{lb} &= \begin{bmatrix}
    H_1^{\min},...,H_{N_V}^{\min}, J_1^{\min},...,J_{N_C}^{\min}, -M, -M, \boldsymbol{0}_{1\times N_{TS}}, \boldsymbol{0}_{1\times N_{TS}}
    \end{bmatrix} \\
    \boldsymbol{ub} &= \begin{bmatrix}
    H_1^{\max},...,H_{N_V}^{\max}, J_1^{\max},...,J_{N_C}^{\max}, M, M, \boldsymbol{1}_{1\times N_{TS}}, \boldsymbol{0}_{1\times N_{TS}}
    \end{bmatrix}
\end{align*}
\end{tcolorbox}
\caption{MILP formulation for \algoN{} method}
\end{BOX}

\section{Results and discussions}\label{sec:rnd}

In this section, an example is presented to show the efficiency of both the algorithms. It is shown by example that the predicted $\eta$ decays and is bounded. Moreover, the \algoN{} method can predict ground states that provide higher bandgap compared to randomly picked ground states. Next, the computational cost of this method is discussed. 

\subsection{Examples}\label{sec:example}

The parametric estimation of Ising model is presented as an application of this method. In this model, the states take a binary form i.e. $N_L = 2$. Traditionally the labels are denoted as $\{+1,-1\}$ and the corresponding energy functions are defined as: 
\begin{gather*}
    U(+1) = +1, \quad U(-1) = -1\\
    V(+1,+1) = V(-1,-1) = 1, \quad V(+1,-1) = -1
\end{gather*}
Therefore, the energy can be effectively written as: 
\begin{gather}\label{eq:Ising}
    E(\boldsymbol{S}) = \sum_{i=1}^{N_V} H_i s_i + \sum_{k=1}^{N_C} J_{k} s_{\pi(k,1)} s_{\pi(k,2)}
\end{gather}

\begin{figure}[tph]
\captionsetup[subfigure]{labelformat=empty}
    \centering
    \begin{subfigure}[p]{0.48\textwidth}
    \centering
    \includegraphics[width=1\linewidth]{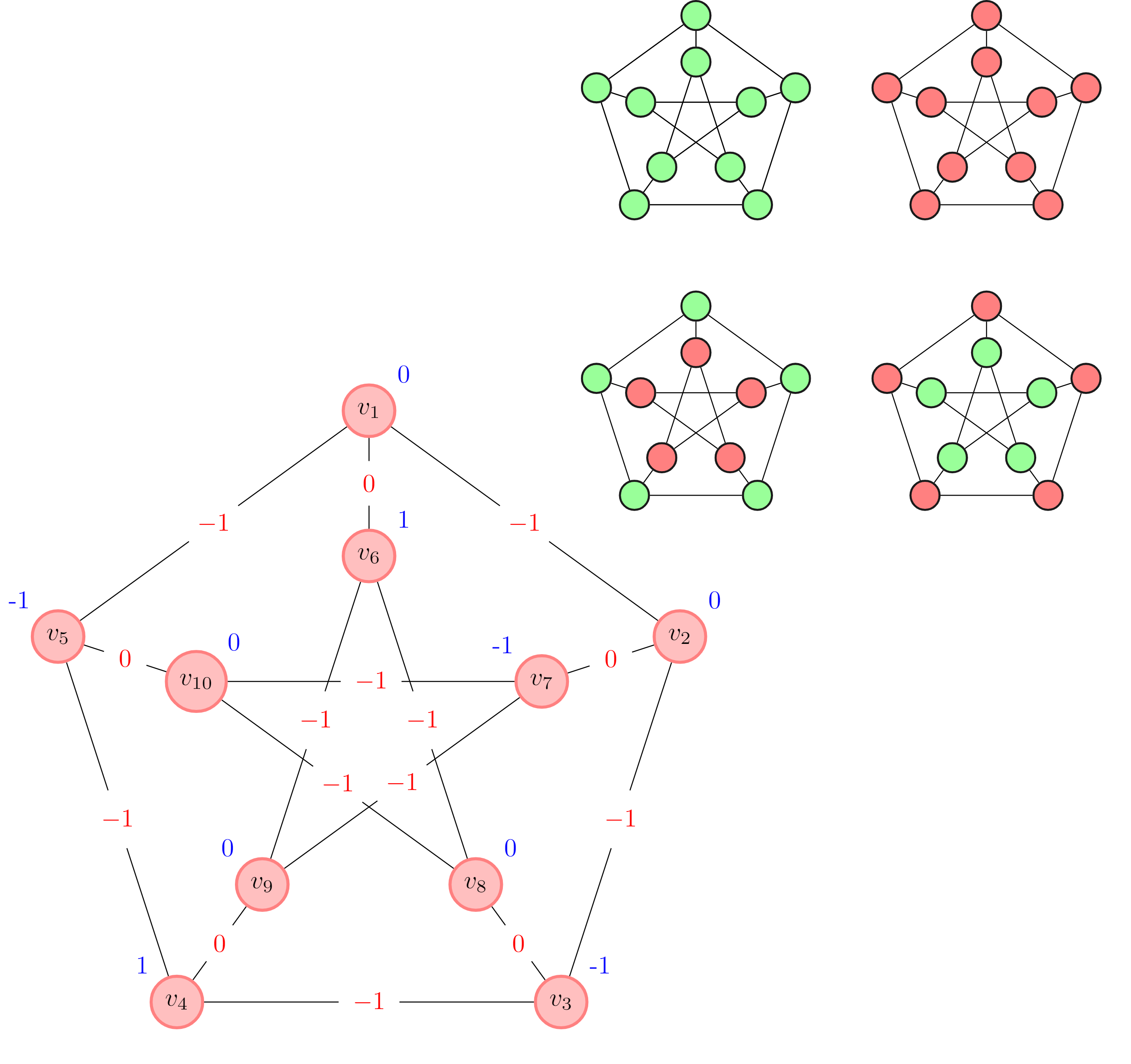}
    \caption{(a)}
    \end{subfigure}\hfill
    \begin{subfigure}[p]{0.48\textwidth}
    \centering
    \includegraphics[width=1\linewidth]{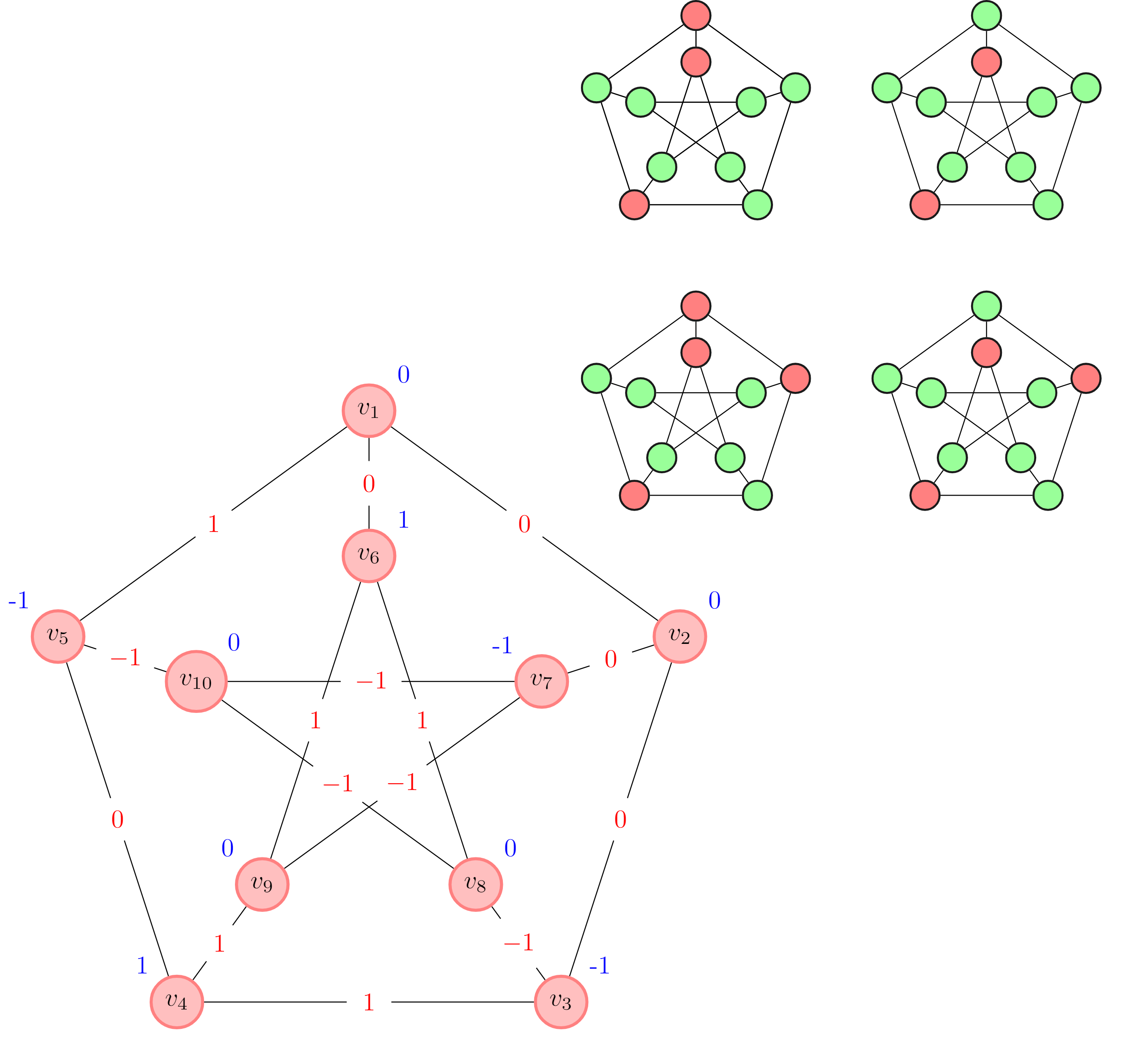}
    \caption{(b)}
    \end{subfigure}\hfill 
    \vspace{5mm}
    \begin{subfigure}[p]{0.7\textwidth}
    \centering
    \includegraphics[width=1\linewidth]{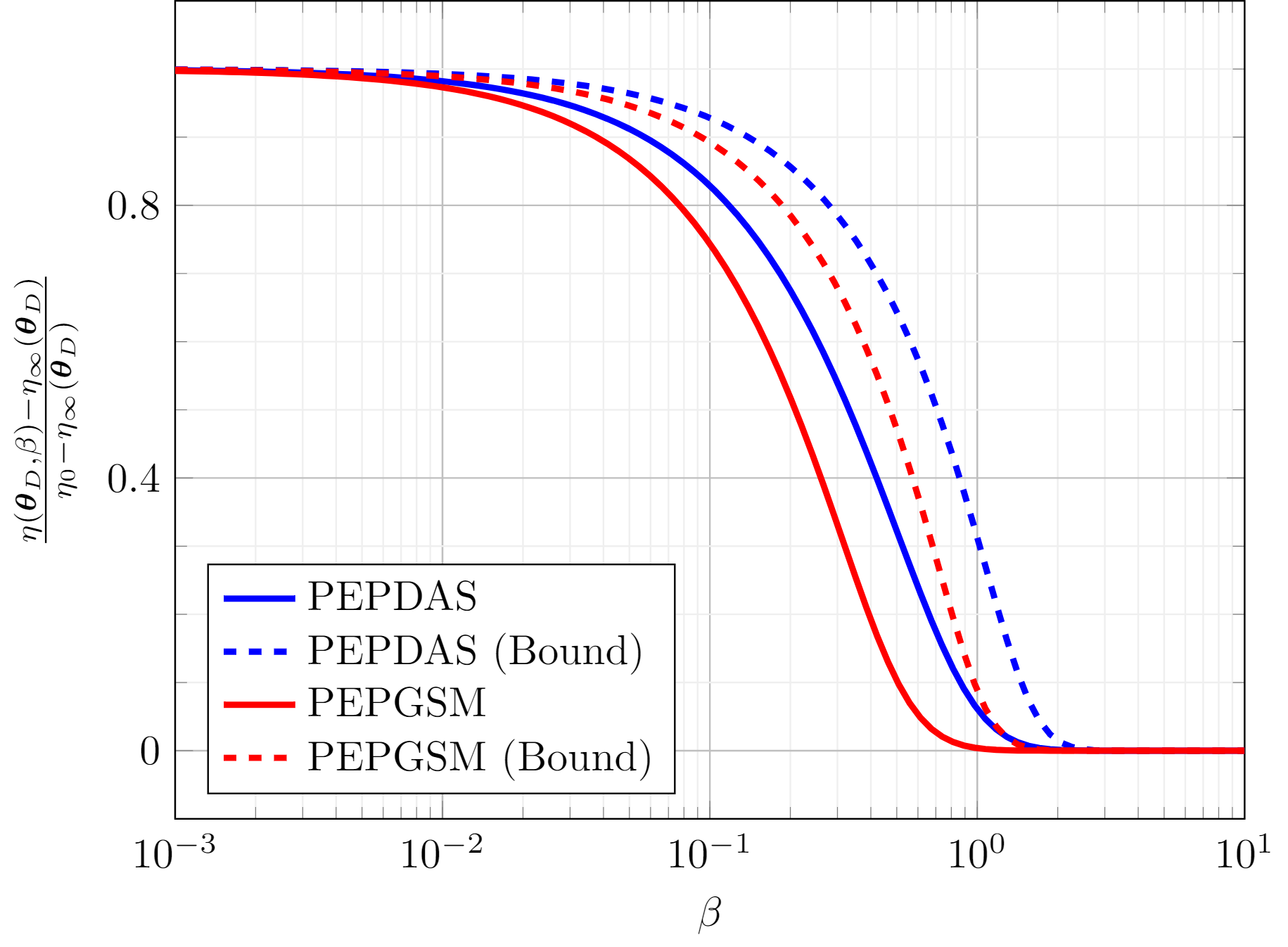}
    \caption{(c)}
    \end{subfigure}
    \caption{Optimal Ising parameters of a Peterson graph with 4 ground states found using (a) \algoG{} method, and (b) \algoN{} method. The ground states are presented as the colored graph in the top right corner of each image. A green label denotes the `$+1$' state, and the red label denotes the `$-1$' state. (c) The normalized Negative log-likelihood of the optimized graphs}
\label{fig:PeteOptim}
\end{figure}

This model is applied on a 10-noded Peterson graph with $|H|\leq 1$ and $|J|\leq 1$. First, the graph is trained by prescribing up to 4 data states using the \algoG{} method. Next, the graph is trained by prescribing the number of states from 1 to 4 using the \algoN{} method. The predicted band gaps are shown in Table\ref{tab:bandGap}. It can be observed that the \algoN{} method predicts the same bandgap as the \algoG{} method for data sets with a size up to 3. However, for 4 data points, the \algoN{} method can identify ground states that provide higher bandgap. The predicted parameters for a graph with four ground states are shown in Fig \ref{fig:PeteOptim}. Likelihood estimates are not well defined in the case of \algoN{} method as it is not trained using the data. However, for comparison, $\eta$ is estimated using the set of ground states in place of the data set. The results for negative log-likelihood of the \algoG{} predicted model and \algoN{} predicted model are shown in Fig\ref{fig:PeteOptim}(c). As expected, \algoN{} predicted model performs better than \algoG{} predicted model in terms of the range of $\beta$ for which they can be used.
The details of the other three models are presented in Appendix \ref{app:Peterson}.

\begin{table}[H]
\centering
\begin{tabular}{@{}|l|c|c|c|c|@{}}
\hline
Algorithm               & $N_{GS}=1$            & $N_{GS}=2$            & $N_{GS}=3$            & $N_{GS}=4$            \\ \midrule
\algoG{}                & 8.0                   & 6.0                   & 4.0                   & 6.0                   \\
\algoN{}                & 8.0                   & 6.0                   & 4.0                   & 4.0                   \\\bottomrule
\end{tabular}
\caption{Predicted maximum band gap for Peterson graph}
\label{tab:bandGap}
\end{table}

\subsection{Computation size}

One of the limiting features of these algorithms is that it grows exponentially with the graph size. An exact number of variables and equations is provided in Table\ref{tab:problemSize}. It should be noted that the number of states, $N_{TS} = N_L^{N_V}$ and is the reason for the large size of the decision variable. The system of equations and inequalities in both algorithms have large sparse blocks which provide some computational easing.
It should also be noted that the sparsity of graph, $G$, does not give considerable advantage in the algorithm as the size of the problem is mainly dictated by the number of labels, $N_T$, and the number of vertices, $N_V$. 

\begin{table}[H]
\centering
\begin{tabular}{@{}|l|c|c|@{}}
\hline
Quantity                        & \algoG{}              & \algoN{}                \\ \midrule
Total variables                 & $N_V+N_C+1+N_{ES}$    & $N_V+N_C+2+2N_{TS}$   \\
Integer (Binary) variables      & $N_{ES}$              & $2N_{TS}$             \\
Inequality conditions           & $2N_{ES}$             & $4N_{TS}+1$           \\ 
Equality conditions             & $N_{DS}$              & $2$                   \\\bottomrule
\end{tabular}
\caption{Variable size for Algorithms \algoG{} and \algoN{}}
\label{tab:problemSize}
\end{table}

\section{Conclusion}\label{sec:conclusion}

Two algorithms were developed and analyzed for estimating parameters of Potts model. The functionality of each method is as follows: 
\begin{enumerate}
\item \algoG{} method estimates the parameters to exactly replicate the ground states as the prescribed data set.
\item \algoN{} method estimates the parameters to identify ground states based on their prescribed quantity.
\end{enumerate}
Both algorithms maximize the band gap between the ground and excited states of the model. It was shown that models optimized in this manner have a higher probability of being in the ground state for a broader range of temperatures. The upper bounds on the optimized model's performance are also estimated. This efficiency is measured in terms of the range of temperature for which ground states' likelihood remains in the desired range. The examples included in the paper show promising practical results on small graphs. As suggested in the main body of the paper, these methods do not scale well with the graph size, and their usage should be restricted to small problems. 

\section{Supplementary Data}
The codes are available at \url{https://github.com/sidsriva/PEP}

\appendix

\section{Proof of theorem}\label{sec:theoremProof}

\textbf{(a)}  Since $\mathcal{S}_{G}(\boldsymbol{\theta},\beta) = \mathcal{S}_D$, the Negative Log Likelhood, $\eta(\boldsymbol{\theta}_D,\beta)$, is estimated as: 
\begin{align*}
    \eta(\boldsymbol{\theta}_D,\beta) = N_{GS} \beta E_0 + N_{GS} \log Z
\end{align*}
The derivative is estimated as:: 
\begin{align}\label{eq:partialEta}
    \frac{d\eta}{d\beta} &= N_{GS} \left((E_{0} - \mathbb{E}(E) \right)
\end{align}
where
\begin{align*}
    \mathbb{E}(E) = \sum_{\boldsymbol{S}\in \mathcal{S}}  E(\boldsymbol{S}) p(\boldsymbol{S}|\boldsymbol{\theta}_D,\beta) 
\end{align*}
Since $\Delta E > 0$, the expected energy is strictly bounded below as $\mathbb{E}(E) > E_0$. Consequently:  
\begin{align*}
    \frac{d\eta}{d\beta} < 0
\end{align*}
In the low temperature limit, Eq\eqref{eq:boltzmann_prob} estimates that the probability of all excited states approaches 0 while all ground states are equally likely with probability $(N_{GS})^{-1}$. Therefore, the value of $\eta$ in this limit is estimated as Eq\eqref{eq:lowTempEta}.

\textbf{(b)}  Let $\boldsymbol{S}_G \in \mathcal{S}_G$ and $P = p(\boldsymbol{S}_G|\boldsymbol{\theta}_D,\beta)$ so that $\eta(\boldsymbol{\theta}_D,\beta) = - N_{GS} \log{P}$. The probability of occurrence of a ground state is given by $N_{GS}P$ and occurrence of a excited state is given as $\left(1-N_{GS}P\right)$. Moreover, for any finite value of $\beta$ both of these probabilities are finite. Therefore, the expectation of energy, $\mathbb{E}$, can be bounded as 
\begin{align*}
    \mathbb{E} = N_{GS} P E_0 + \sum_{\boldsymbol{S}\in \mathcal{S}_E}  E(\boldsymbol{S}) p(\boldsymbol{S}|\boldsymbol{\theta}_D,\beta) \leq N_{GS} P E_0 + (1- N_{GS} P) E_1
\end{align*}
Substituting in Eq\eqref{eq:partialEta},  
\begin{align*}
    \frac{d \eta}{d\beta} = E_{0} - \mathbb{E}(E) & \leq  \left(  N_{GS}P - 1\right) N_{GS} \Delta E
\end{align*}
Substituting $P = e^{-\eta/N_{GS}}$ gives the following differential inequality 
\begin{align}
    \frac{d \eta}{d\beta} \leq  \left(  N_{GS} e^{-\eta/N_{GS}} - 1\right) N_{GS} \Delta E
\end{align}
Consider the differential equation for $\beta\in[0,\infty)$, 
\begin{align}
    \frac{d \xi }{d\beta} = \left(  N_{GS} e^{-\xi /N_{GS}} - 1\right) N_{GS} \Delta E
\end{align}
with initial condition $\xi(\boldsymbol{\theta}_D,0) = \eta(\boldsymbol{\theta}_D,0) = N_{GS}\log N_{TS} $. Noting that $N_{GS} e^{-\xi /N_{GS}} - 1  = N_{GS}P-1 > 0$, this ODE is integrated to give the following solution: 
\begin{align}
    \xi(\boldsymbol{\theta}_D,\beta) = N_{GS} \log{ \left( N_{GS} + N_{ES} e^{-\beta \Delta E} \right) }
\end{align}
Using Comparison Lemma \cite{khalil_2002}, for all $0<\beta<\infty$, 
\begin{align}
    \eta(\boldsymbol{\theta}_D,\beta) \leq \xi(\boldsymbol{\theta}_D,\beta)
\end{align}
This proves the upper bound. The lower bound is a direct consequence from monotonicity proved in part 1. 

\textbf{(c)}  For any $\beta < \infty$ 
\begin{align*}
    \eta(\boldsymbol{\theta}_D,\beta)  - N_{GS} \log{N_{GS}} \leq N_{GS} \log\left( 1+ \frac{N_{ES}}{N_{GS}} e^{-\beta \Delta E} \right)
\end{align*}
For any $\epsilon>0$, choose a $\beta > \beta^*(\epsilon)$ using Eq\eqref{eq:betaMatch}  and observe that, 
\begin{align*}
    N_{GS}\log\left( 1+ \frac{N_{ES}}{N_{GS}} e^{-\beta \Delta E} \right) < \epsilon
\end{align*}
This proves the third statement. 

\section{Optimized Graphs}

\subsection{K-3 graph}\label{app:K3}
\begin{figure}[tph]
\captionsetup[subfigure]{labelformat=empty}
    \centering
    \begin{subfigure}[p]{0.32\textwidth}
    \centering
    \includegraphics[width=1\linewidth]{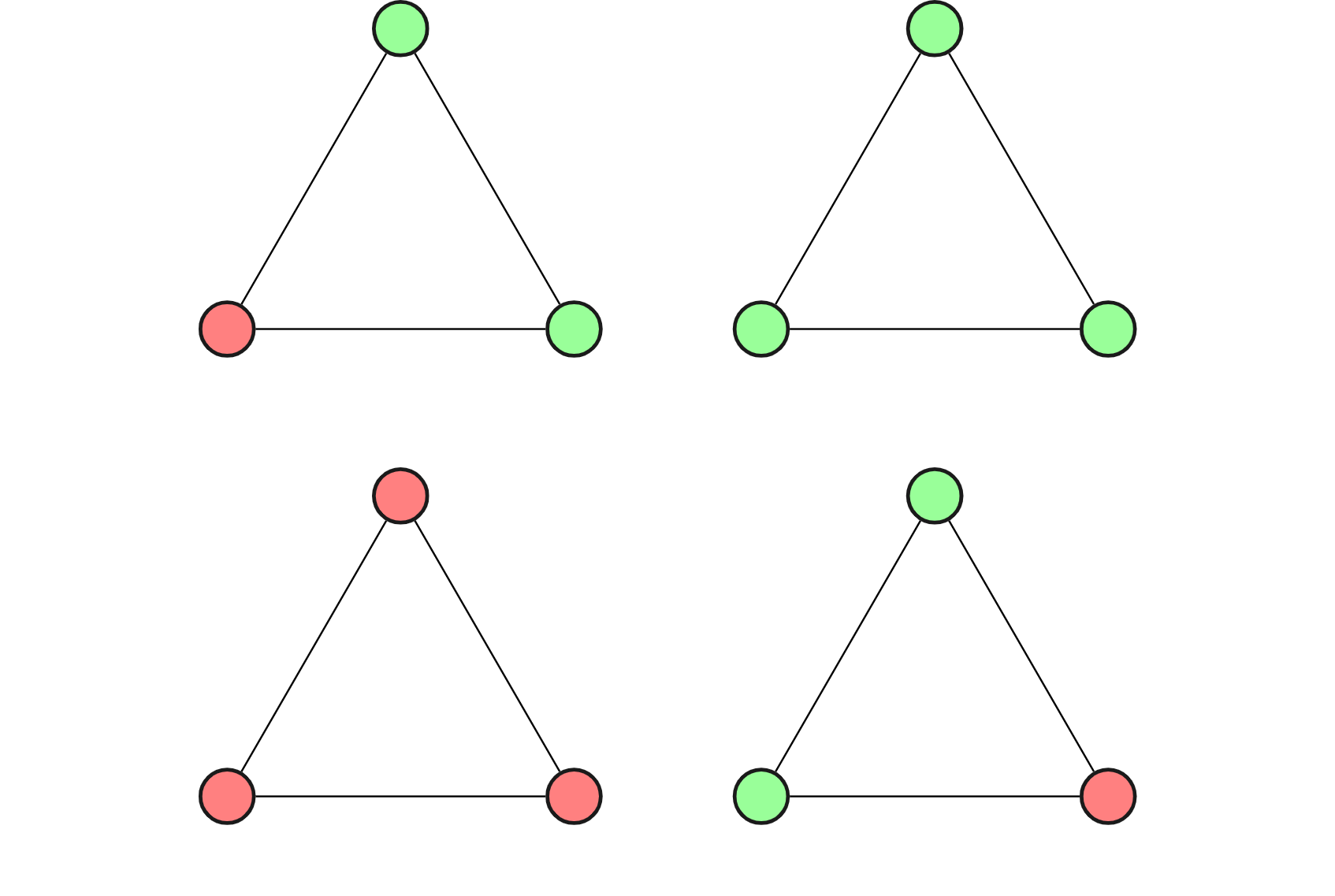}
    \caption{(a)}
    \end{subfigure}\hfill
    \begin{subfigure}[p]{0.32\textwidth}
    \centering
    \includegraphics[width=1\linewidth]{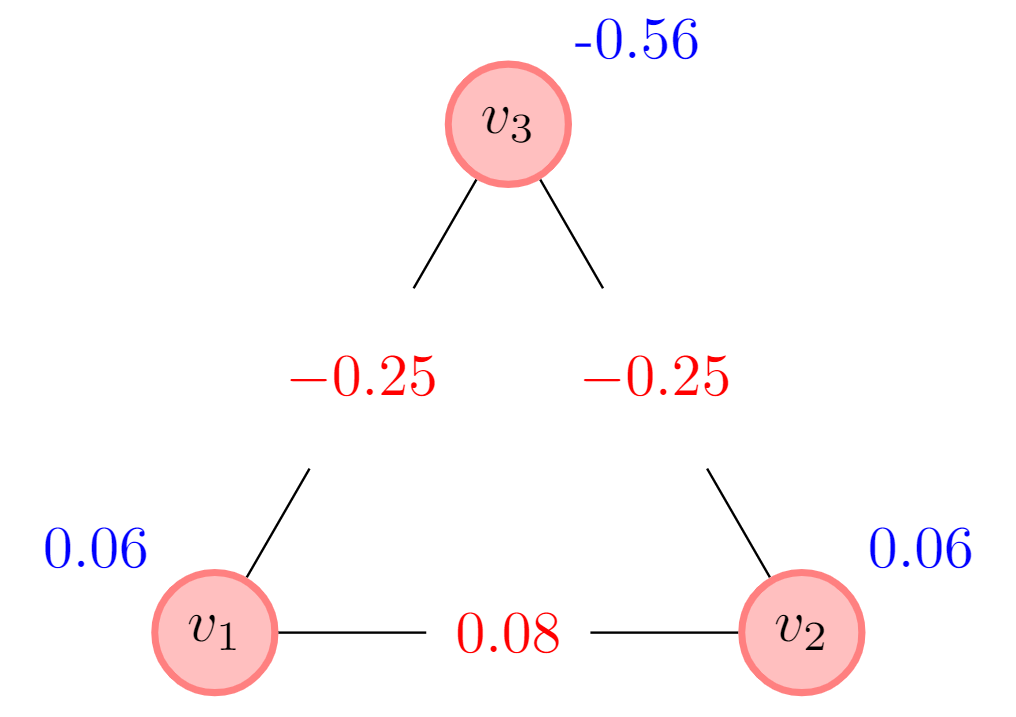}
    \caption{(b)}
    \end{subfigure}\hfill 
    \vspace{5mm}
    \begin{subfigure}[p]{0.32\textwidth}
    \centering
    \includegraphics[width=1\linewidth]{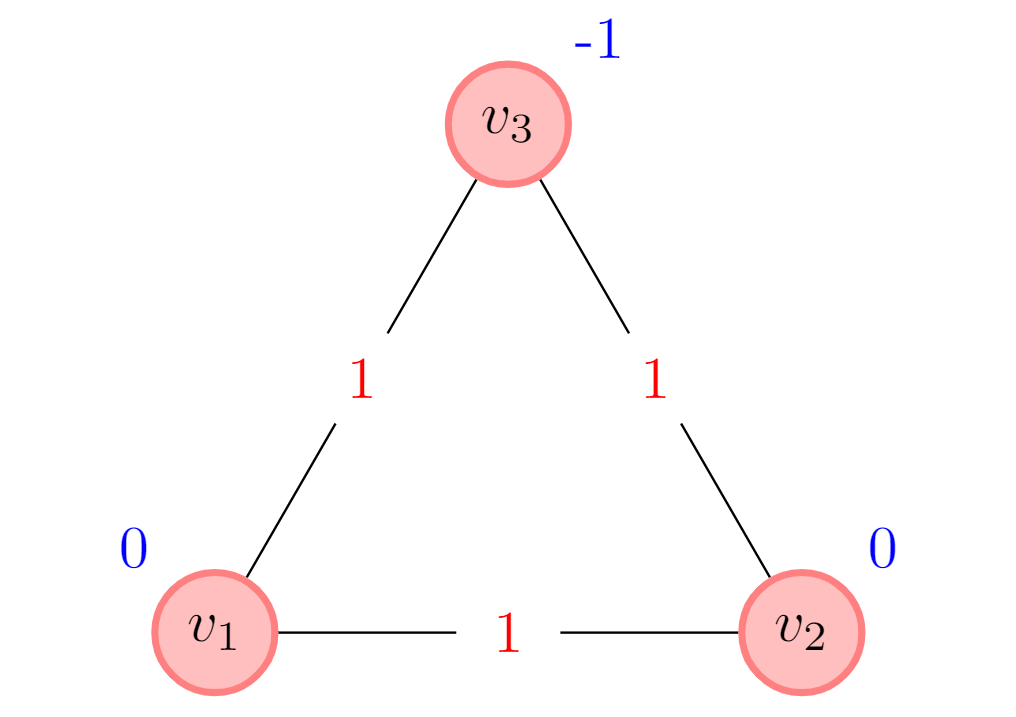}
    \caption{(c)}
    \end{subfigure}
    \caption{(a) Training data set of states with green representing a `+1' state and red representing a `-1' state. (b) Optimized graph using minimization of Negative Log-likelhood at $\beta=1$ (c) Optimized graph using \algoG{} method. The field terms are mentioned in blue color and interaction terms are mentioned in red color }
\label{fig:OrGate}
\end{figure}

A fully connected 3-noded graph is optimized for 4 data states. The energy of the graph is modeled using Ising model Eq\eqref{eq:Ising} with $|H|\leq 1$ and $|J|\leq 1$. The optimized parameters using the (1) Minimization of Negative Log-likelihood, and (2) \algoG{} method are presented in Fig.\ref{fig:OrGate}

\subsection{Peterson graph}\label{app:Peterson}

\begin{figure}[tph]
\centering

\includegraphics[width=0.45\linewidth]{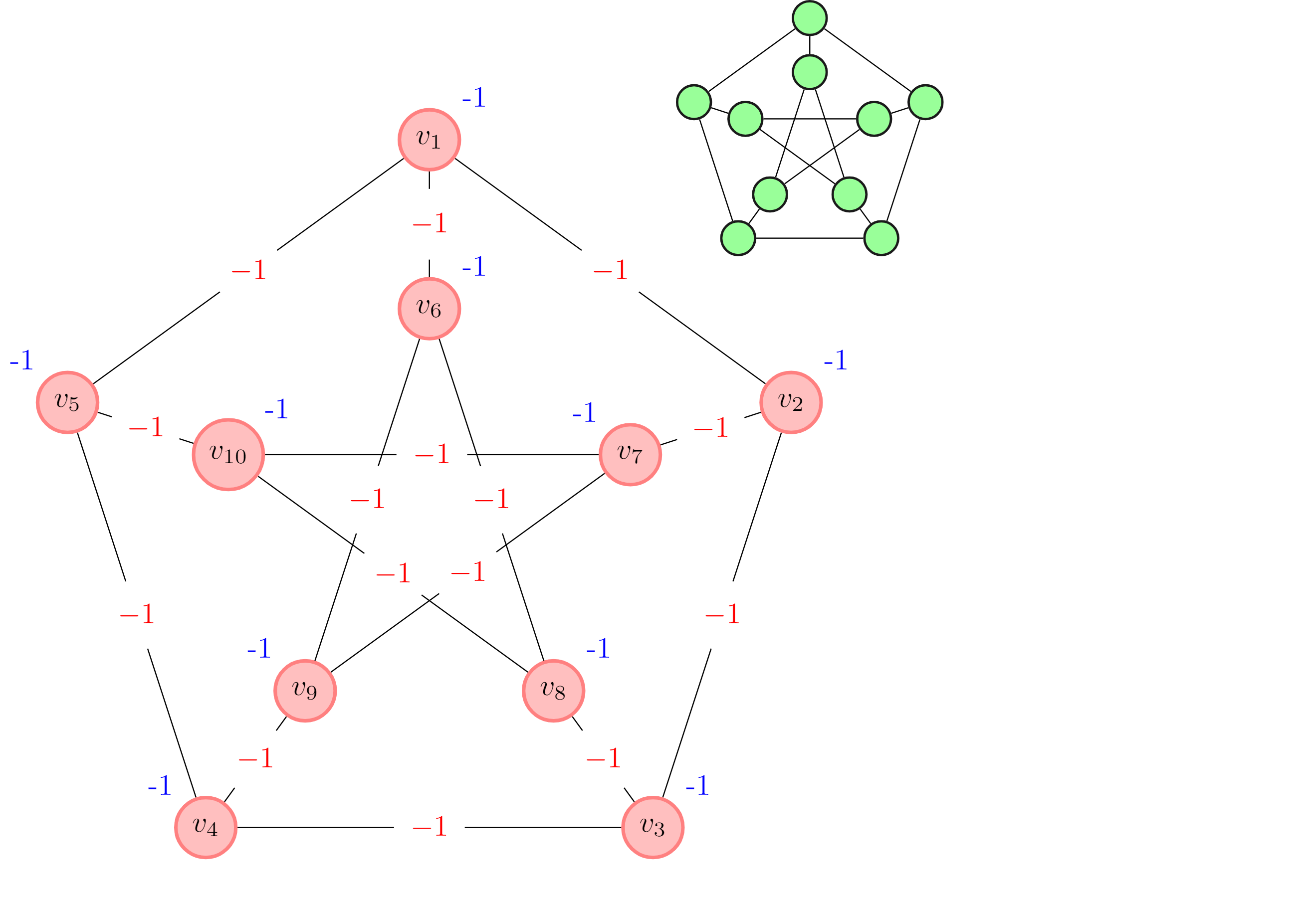}
\includegraphics[width=0.45\linewidth]{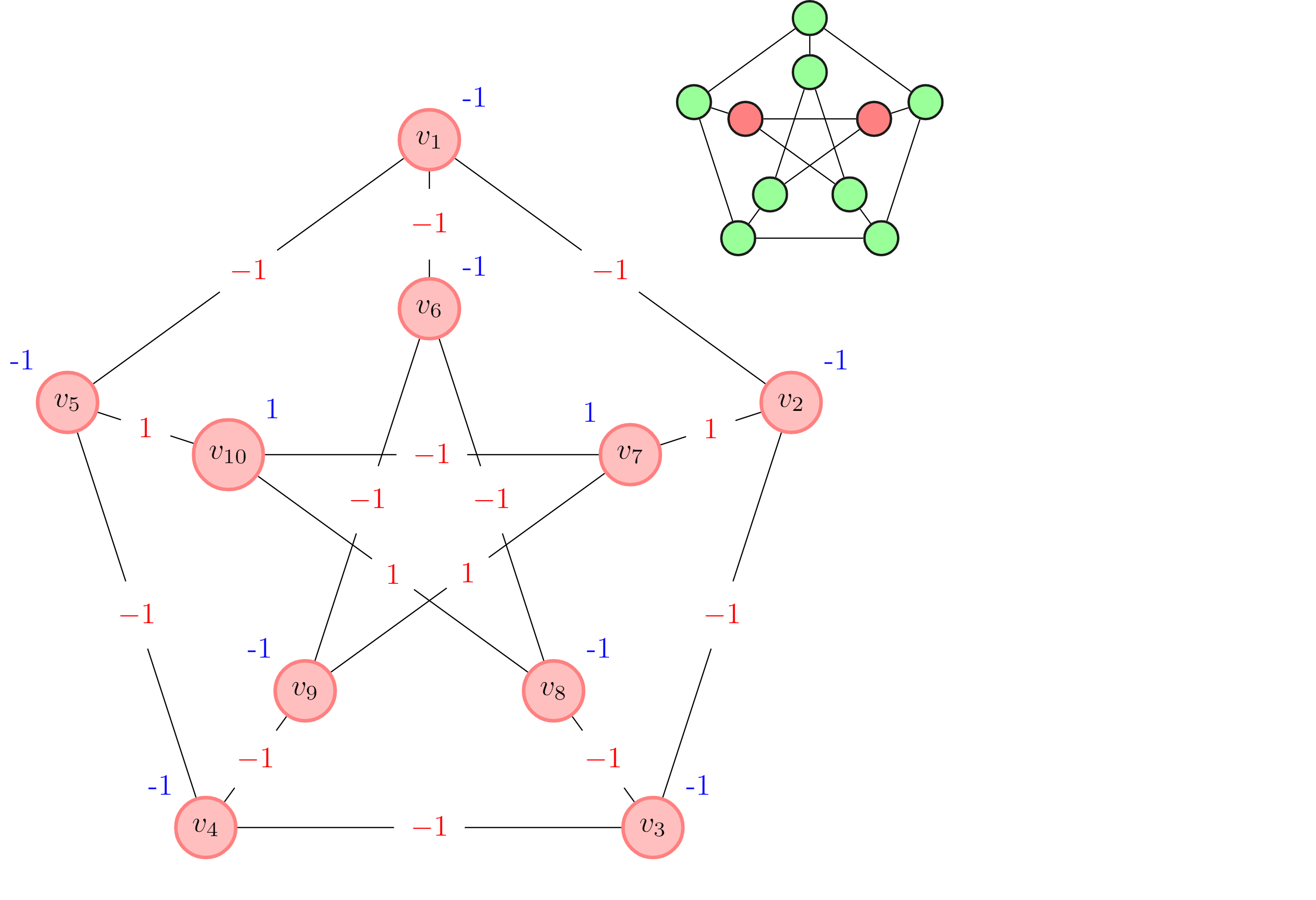}

\includegraphics[width=0.45\linewidth]{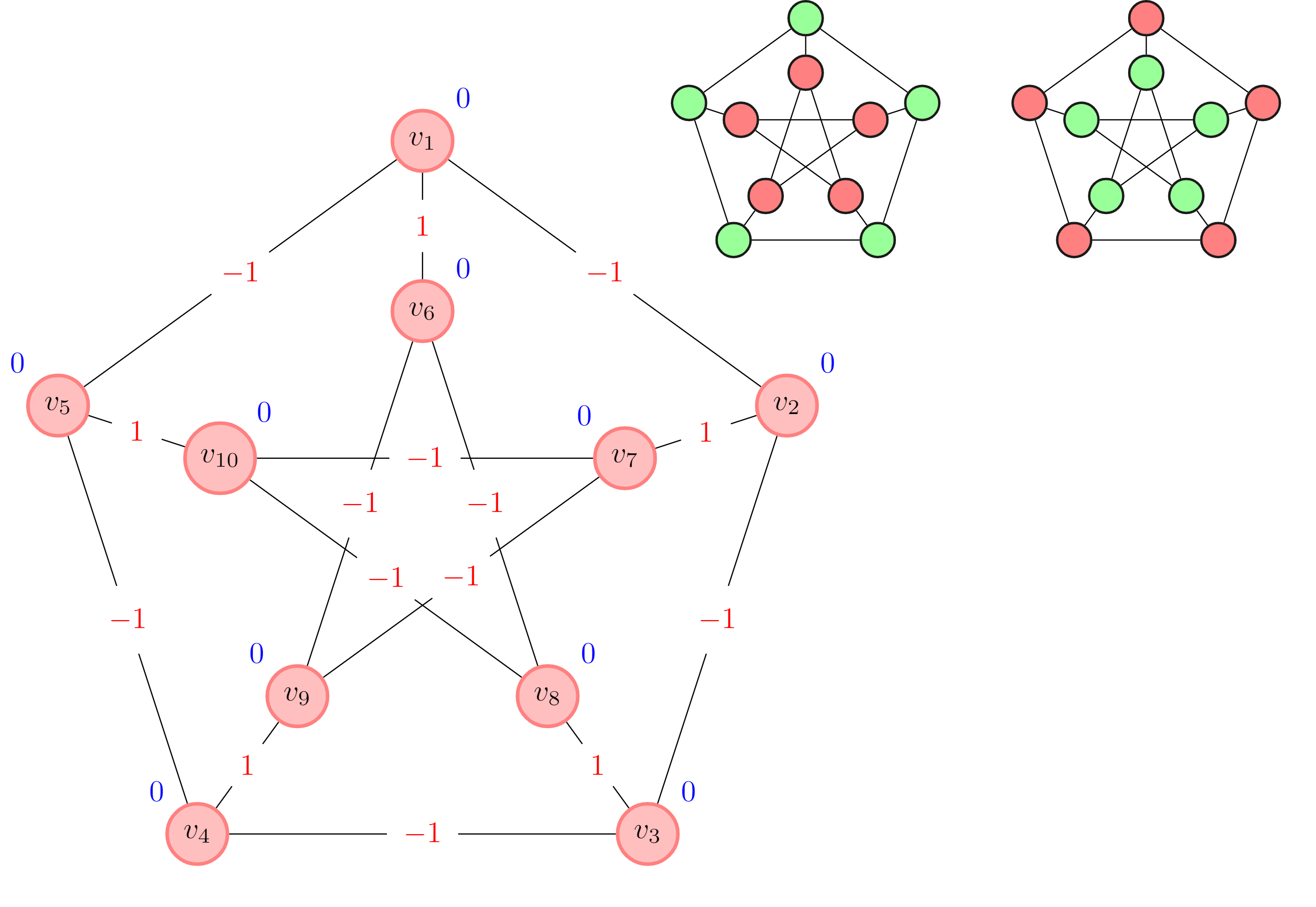}
\includegraphics[width=0.45\linewidth]{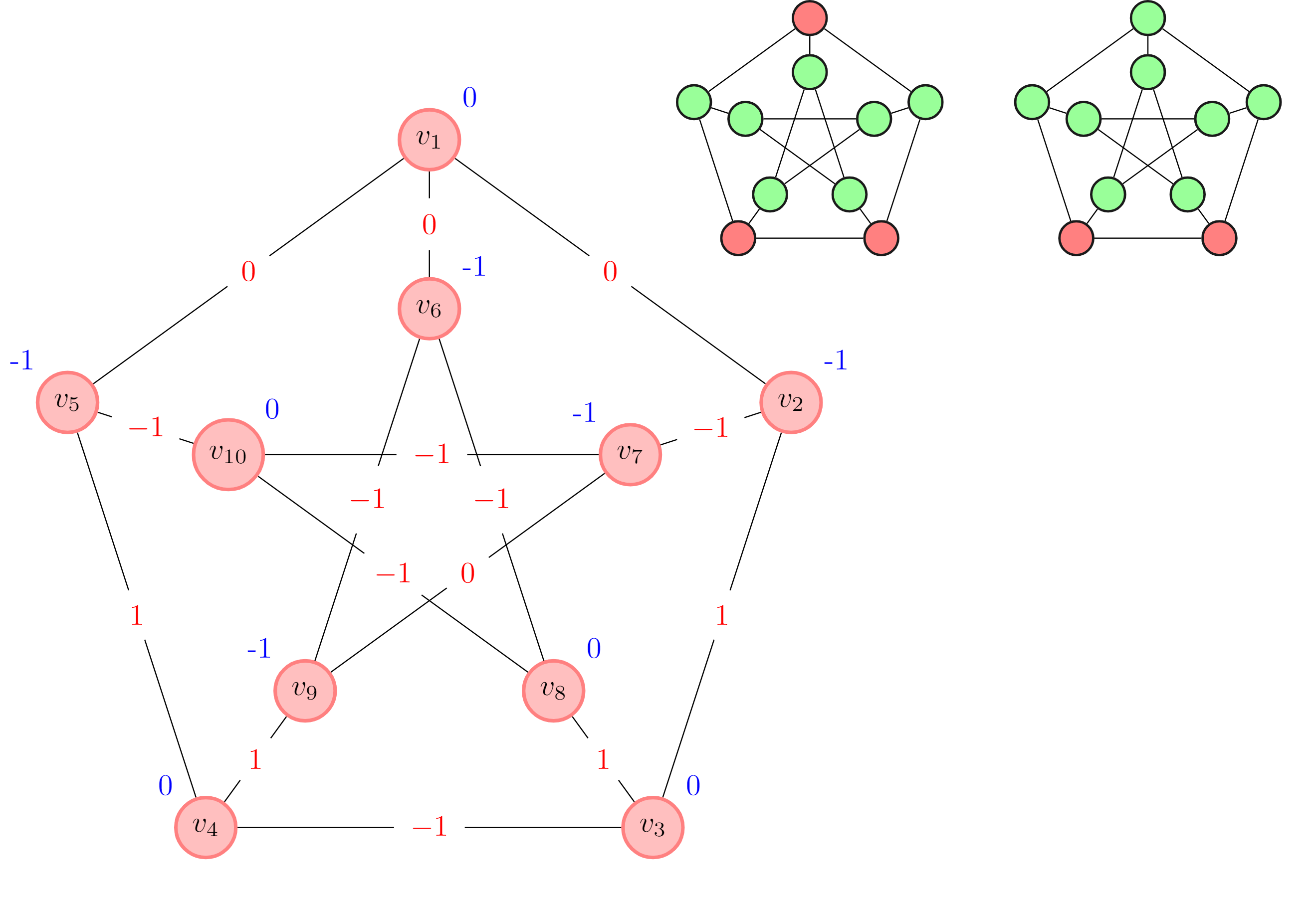}

\includegraphics[width=0.45\linewidth]{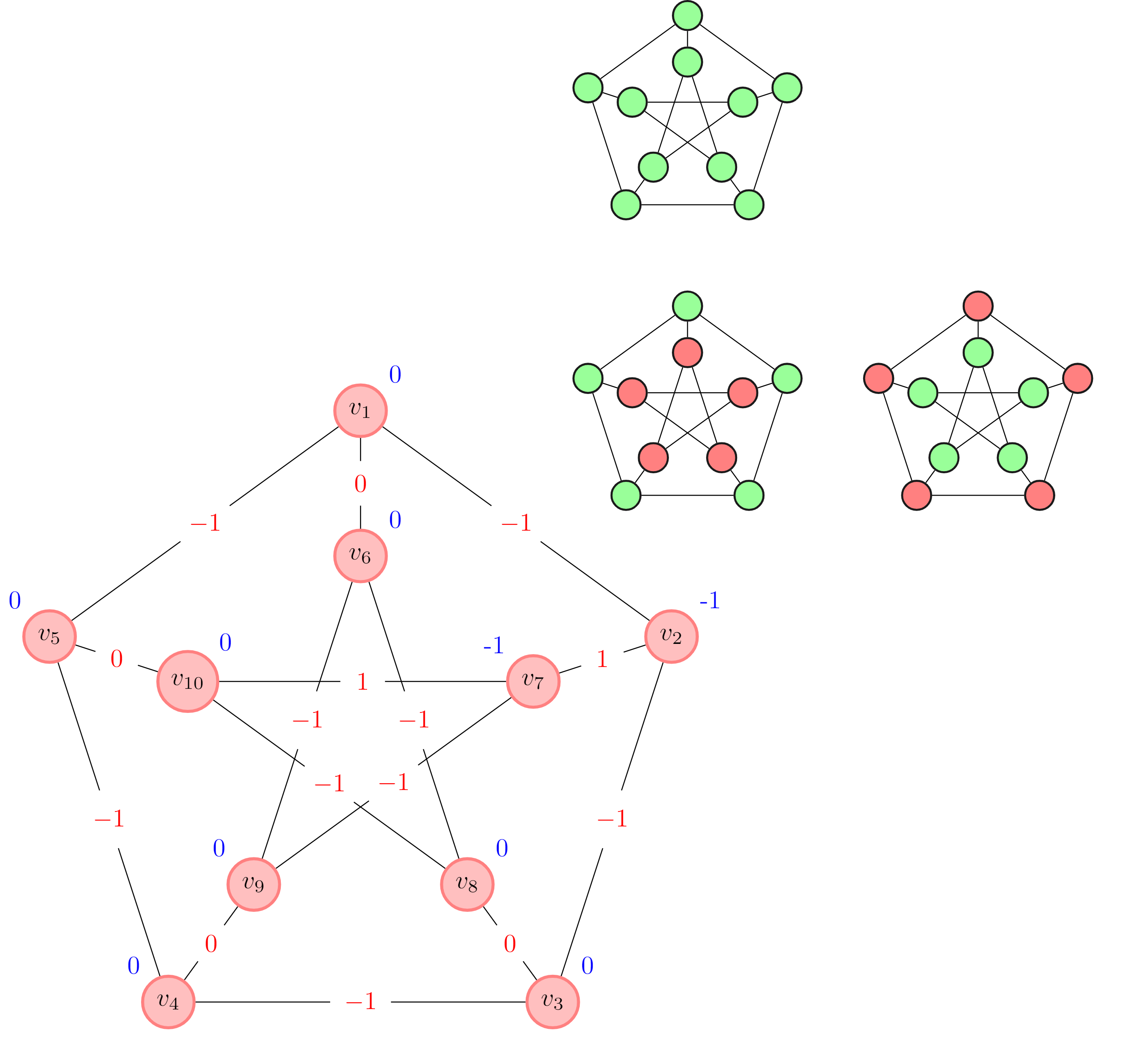}
\includegraphics[width=0.45\linewidth]{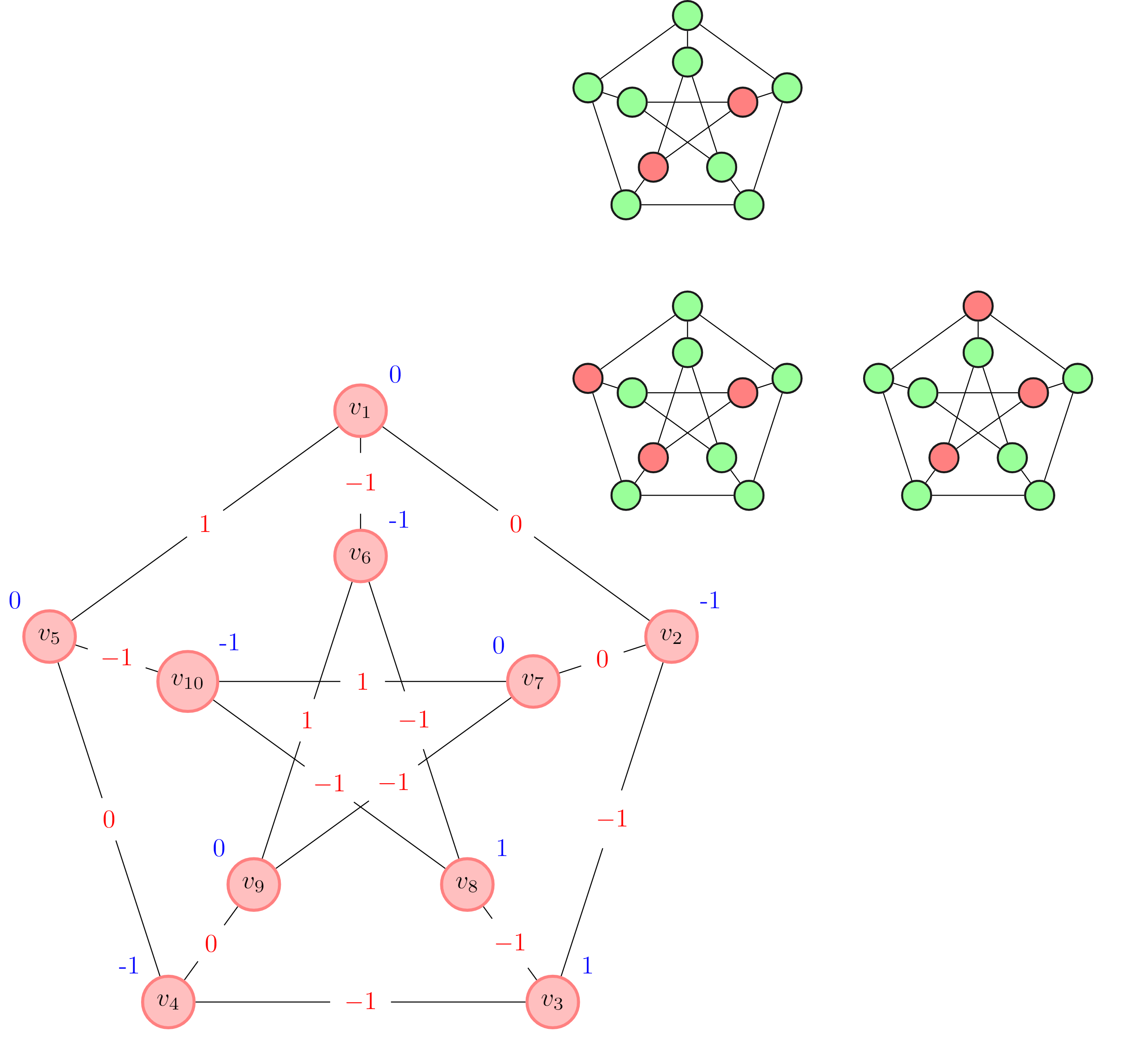}

\caption{Optimal Ising parameters of a Peterson graph found using \algoG{} method (left) and \algoN{} method (right). The ground states are presented as the colored graph in the top right corner of each image. A green label denotes the `$+1$' state and the red label denotes the `$-1$' state.}
\label{fig:PeteOptim3}
\end{figure}

\begin{figure}[tph]
\centering
\centering
\includegraphics[width=0.8\linewidth]{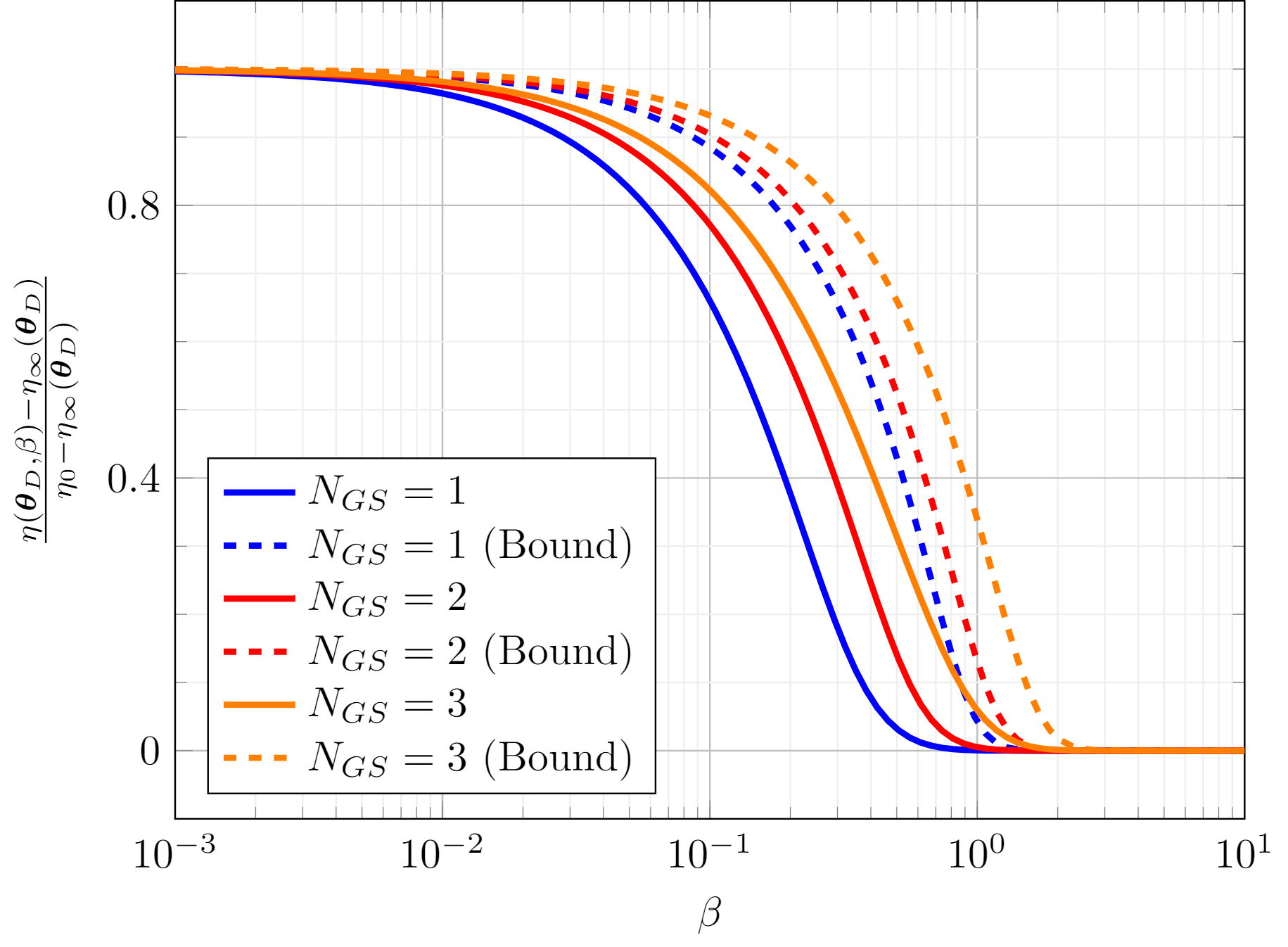}
\caption{Normalized Negative log likelihood and their respective bounds for Peterson graphs trained using \algoG{} method}
\label{fig:PeteNLL3}
\end{figure}

A Peterson graph is first optimized for upto 3 user prescribed data states using \algoG{} method. Then it is optimized for 3 ground states using \algoN{} method. The energy of the graph is modeled using Ising model Eq\eqref{eq:Ising} with $|H|<1$ and $|J|<1$. The optimized graphs are presented in Fig\ref{fig:PeteOptim3} and their respective Negative log likelhood is presented in Fig.\ref{fig:PeteNLL3}.

\bibliographystyle{unsrt}  
\bibliography{references}  

\end{document}